\documentclass[11pt]{article}
\usepackage{amsmath}
\usepackage{amssymb}
\usepackage{theorem}
\usepackage[numbers,sort&compress]{natbib}
\newtheorem{proposition}{Proposition}[section]

\newtheorem{definition}{Definition}[section]
\newtheorem{theorem}{Theorem}[section]

\newtheorem{lemma}{Lemma}[section]
\newtheorem{remark}{Remark}[section]
\newtheorem{example}{Example}[section]
\newtheorem{corollary}{Corollary}[section]
\newcommand{\N}{\mathbb{N}}
\newcommand{\Z}{\mathbb{Z}}
\renewcommand{\P}{\underline{P}}

\newcommand{\C}{\underline{C}}
\newcommand{\A}{\underline{A}}
\def\petitcarre{\vrule height4pt width 4pt depth0pt}
\def\enddim{\relax\ifmmode\eqno{\hbox{\petitcarre}}
\else
{\unskip\nobreak\hfil\penalty50
   \hskip2em\hbox{}\nobreak\hfil
   \petitcarre
   \parfillskip=0pt \finalhyphendemerits=0
  \par\medskip}\fi}

\def \begdim {\noindent {\sc Proof} : \par \noindent}

\DeclareMathOperator{\Card}{Card}
\DeclareMathOperator{\supp}{supp}

\begin{document}

\numberwithin{theorem}{section}
\numberwithin{equation}{section}
\numberwithin{figure}{section}
\numberwithin{table}{section}

\title{\Large \bf
Some conjectures on codes
\thanks{
Partially supported by
the $FARB$ Projects {\it ``Aspetti matematici e applicativi nella teoria dei codici
e linguaggi formali''}
(University of Salerno, 2015) and {\it
``Linguaggi formali e codici: metodi combinatori e orientamenti
applicativi''} (University of Salerno, 2016).}}

\author{Clelia De Felice, \\
Universit\`a degli Studi di Salerno, \\
defelice@dia.unisa.it}
\maketitle

%------------
\begin{abstract}
Variable-length codes are the {\it bases} of the free
submonoids of a free monoid. There are some important longstanding open questions
about the structure of finite {\it maximal codes}.
In this paper we discuss this conjectures and their relations with factorizations of
cyclic groups.
\end{abstract}
%----------------------

\thispagestyle{plain}

%------------------------------
\section{Introduction}

The theory of {\it variable-length codes} takes its origin
in the framework of the theory of information, since Shannon's early
works in the 1950's. An algebraic theory of codes was subsequently initiated
by Sch\"{u}tzenberger,
who proposed in \cite{SC55} the semigroup theory as a
mathematical setting for the
study of these objects.
In this context the theory of codes has been extensively
developed, showing strong relations with automata theory, combinatorics
on words, formal languages and the theory of semigroups (see \cite{BPR}
for a complete treatment of this topic and \cite{BDFDLPRR1,BDFDLPRR2,BDFDLPRR3,BDFDLPRR4,BDFPRR}
for recent results on strong connections between codes, combinatorics on words and free groups).
In this paper
we follow this algebraic approach and codes
are defined as the {\it bases} of the free
submonoids of a free monoid.

We are interested in some important longstanding open questions
about the structure of finite {\it maximal codes}
(maximal objects in the class of codes for the order of set inclusion).
One of these conjectures asks whether
any finite maximal code $C$ is {\it positively factorizing} \cite{SC}, that is if
there always exist
finite subsets $P$, $S$ of $A^*$ such that
\begin{eqnarray} \label{FC}
\underline{C} - 1 & = & \underline{P}(\underline{A} - 1)\underline{S}
\end{eqnarray}
(here $1$ is the empty word and \underline{X} denotes
the {\it characteristic polynomial}
of a finite language $X$, i.e., the formal sum of its elements).

The above conjecture
was formulated by
Sch\"{u}tzenberger but, as
far as we know, it does not appear
explicitly in any of his papers.
It was quoted as the
{\it factorization conjecture} in \cite{P1}
for the first time and then also reported
in \cite{BPR,BER}.
The major contribution to this conjecture is due to
Reutenauer \cite{REU83,REU85}. In particular,
he proved that for any finite
maximal code $C$ over $A$, there exist polynomials
$P, S \in \Z \langle A \rangle$ such that
$\underline{C} - 1 = P(\underline{A} - 1)S$.
Other partial results concerning this conjecture may be found in
\cite{BO78,BO79,BO81,DF93,DF13,DFReu,RES,ZHG}.

The conjecture is still open and weaker forms of it have been proposed and reported below.

Two words $x,y$ are {\it commutatively equivalent} if
the symbols of $y$ can be reordered to make $x$.
Two sets $X,Y$ are commutatively equivalent if there is a bijection $\phi$ from
$X$ onto $Y$ such that for every $x \in X$, $x$ and $\phi(x)$ are commutative equivalent.
A well known class of codes is that
of {\it prefix} codes, i.e., codes such that
none of their words is a left factor of another.
A code $X \subseteq A^*$ is {\it commutatively prefix} if there exists a prefix
code $Y \subseteq A^*$ which is commutatively equivalent to $X$.

It is conjectured that every finite maximal code is commutatively prefix.
This is the {\it commutative equivalence conjecture},
due to Perrin and Sch\"{u}tzenberger and inspired by a problem of information theory \cite{PS77}.
Any positively factorizing code is commutatively prefix.
Partial results on the commutative equivalence conjecture have been
proved in \cite{MR,PS77} and a formulation of it, when restricted to a two-letter alphabet,
in terms of continued fractions of a finite length has been given
in \cite{DL}.

A third conjecture takes into account {\it bayonet} codes, i.e., codes such that each of its words has the form $a^iba^j$,
$a,b \in A$. It is conjectured that for any finite bayonet code $X$ which can be embedded
in a finite maximal code, one has $\Card(X) \leq \max\{|x| ~|~ x \in X \}$.
This is the {\it triangle conjecture}, due to Perrin and Sch\"{u}tzemberger \cite{PS81}.
If $X$ is a finite maximal code and $X$ is commutatively prefix, then $X \cap a^*ba^*$
verifies the triangle conjecture, for any $a,b \in A$.
Partial results on the triangle conjecture have been proved in \cite{DF83,HAN,PS}.

Originally the three conjectures were proposed for codes with
no additional hypothesis. In 1985 Shor found a bayonet code $X$ such that
$\Card(X) > \max\{|x| ~|~ x \in X \}$ \cite{Shor}. Thus the conjectures were restricted
as above to the smaller class of finite maximal codes and its subsets.

Notice that there are finite codes which are not contained in any finite maximal code
\cite{RES}. The {\it inclusion problem}, for a finite code $X$, is the existence of a finite maximal
code containing $X$. The {\it inclusion conjecture} claims that the inclusion problem
is decidable.

In this paper we focus on some relations between {\it factorizations of cyclic groups}, positively
factorizing codes and finite maximal codes.
We recall that a pair $(T,R)$ of subsets of $\N$ is a factorization of
$\Z_n$ if for any $z \in \{0, \ldots, n-1 \}$, there exists a unique pair $(t,r)$,
with $t \in T$ and $r \in R$, such that $t + r = z \pmod{n}$ \cite{SZSA}.

Known results linking factorizations and
positively factorizing codes are reported in Section \ref{FPF}. Recent results concerning relations
between factorizations of cyclic groups and finite maximal codes, proved in \cite{ZHSH}, will be described in Section
\ref{FMC} (see \cite{BPR,DFRes,LAM97,RSS,SZ} for former results on these relations). Finally, we discuss connections
between the former and the latter results in Section \ref{MCGA} and
some issues that follow in Section \ref{OP}.

%---------------------------------------------

\section{Basics} \label{Ba}

\subsection{Codes and words}

Let $A^{*}$ be the {\it free monoid}
generated by a finite alphabet $A$
and let $A^+=A^{*} \setminus 1$ where $1$ is
the empty word.
For a word $w \in A^*$ and a
letter $a \in A$, we denote by $|w|$ the {\it length}
of $w$ and by $|w|_a$ the number of the occurrences
of $a$ in $w$.

A {\it code} $C$ is a subset
of $A^{*}$ such that, for all $h, k \geq 0$ and
$c_1, \ldots , c_h, c'_1, \ldots , c'_k \in C$, we have
\[
c_1 \cdots c_h= c'_1 \cdots c'_k \quad \Rightarrow \quad
h=k \quad \mbox{and} \quad c_i=c'_i
\quad \mbox{for} \quad i = 1, \ldots , h.
\]
A set $C \subseteq A^+$, such that $C \cap CA^+ = \emptyset$,
is a {\it prefix} code. $C$ is a {\it suffix} code if
$C \cap A^+C = \emptyset$ and $C$ is a {\it bifix}
code when
$C$ is both a suffix and a prefix code.
A code $C$ is a {\it maximal} code over $A$ if for each code $C'$
over $A$ such that $C \subseteq C'$ we have $C=C'$.
If $C$ is a finite maximal code, for each letter $a \in A$, there is an integer
$n \in \N$ such that $a^n \in C$, called the {\it order} of $a$ relative to $C$.

\subsection{Polynomials}

Let $\Z \langle A \rangle$
(resp. $\N \langle A \rangle$)
denote the semiring of the {\it
polynomials} with noncommutative variables in $A$
and integer (resp. nonnegative integer)
coefficients.
For a finite subset
$X$ of $A^{*}$,  $\underline{X}$ denotes its
{\it characteristic polynomial}, defined by
$\underline{X}= \sum_{x \in X} x$.
Therefore, ``characteristic polynomial'' will be synonymous
with ``polynomial with coefficients $0,1$''.
For a polynomial
$P$ and a word $w \in A^{*}$, $(P,w)$
denotes the coefficient of $w$ in $P$ and
we set $\supp(P) = \{ w \in A^* ~|~ (P,w) \not= 0\}$.
If $\supp(P) = \emptyset$, then $P = 0$ is the null polynomial.
When we write $P \geq Q$, with $P,Q \in \Z \langle A \rangle$,
we mean that $(P,w) \geq (Q, w)$, for any $w \in A^*$.
In particular, $P \geq 0$ means that
$P \in \N \langle A \rangle$.
For $P \in \Z \langle A \rangle$, $b \in A$ and
$g \in \N$, we
denote by $P_g$ polynomials such that
$$\forall w \in A^* \quad
(P_g, w) = \begin{cases} (P, w) & \text{if $|w|_b = g$}, \\
0 & \text{otherwise.} \end{cases}$$
We write, as usual, $\Z[a]$ and $\N[a]$ instead of
$\Z \langle a \rangle$  and $\N \langle a \rangle$.
The map which associates the polynomial $\sum_{n \in \N} (H,n)a^{n} \in \N[a]$
to a finite multiset $H$ of
nonnegative integers, is a bijection between
the set of the finite multisets $H$ of nonnegative integers
and $\N[a]$.
We represent this bijection by the notation
$a^H = \sum_{n \in \N} (H,n)a^{n}$.
For example, $a^{\{0, 0, 1, 1, 1, 3, 4 \}} = 2 + 3a + a^3 + a^4$.
Consequently, the following computation rules are defined:
$a^{M+L}=a^{M}a^{L}$,
$a^{M \cup L}= a^{M} + a^{L}$,
$a^{\emptyset}=0$, $a^0=1$.

\subsection{Positively factorizing codes}

Given a finite maximal code $C$, a {\it factorization}
$(P,S)$ for $C$ is a pair of polynomials
$P,S \in \Z \langle A \rangle$
such that $\C=P(\A-1)S+1$.
The following result shows that any finite maximal code has a factorization.

\begin{theorem} \label{ine} \cite{REU85}
Let $C \in \N \langle A \rangle$,
with $(C,1)=0$, and let
$P,S \in \Z  \langle A \rangle$ be
such that $C=P(\A-1)S+1$. Then, $C$ is the characteristic polynomial
of a finite maximal code. Furthermore, if
$P,S \in \N \langle A \rangle$,
then $P,S$ are polynomials with coefficients $0,1$.
Conversely, for any finite maximal code $C$
there exist
$P,S \in \Z \langle A \rangle$
such that $\C=P(\A-1)S+1$.
\end{theorem}

Of course,
$(P,S)$ is a factorization for $C$ if and
only if the same holds for $(-P, -S)$
We say that a factorization $(P,S)$ for $C$ is
{\it positive} if
$P,S$ or $-P, -S$ have coefficients $0,1$.\footnote[1]{Note that in this paper we use
the term ``positive factorization'' with a slightly different
meaning with respect to the definition of the same term
in \cite{BPR}.}
Any code $C$ having a positive factorization
is finite, maximal
and is called a {\it (positively) factorizing code}.

Finite maximal prefix codes are the simplest
examples of positively factorizing codes.
Indeed, $C$ is a finite maximal prefix code if and
only if $\C = \P(\A-1) + 1$ for a finite subset
$P$ of $A^*$ \cite{BPR}.
In the previous relation,
$P$ is the set of the proper prefixes
of the words in $C$.

Let $C$ be a finite maximal code over $A$, let $a$ be a letter and
let $n$ be its order.
Assume that $(P,S)$ is a factorization for $C$
and $P,S$ have coefficients $0,1$.
Then $P,S \in \Z \langle A \rangle$
are such that $\C=P(\A-1)S+1$.
Thus,
there exists
$(I,J) \subseteq \N$
and for all $b \in A \setminus \{ a \}$, finite sets $I'$, $J'$, $L_{i}$,
$M_{j}$ of nonnegative integers,
such that
\begin{eqnarray} \label{EC1}
C_0 &=& a^n, \; P_0=a^I,\; S_0=a^J,\; a^Ia^J=\frac{a^{n}-1}{a-1},
\end{eqnarray}
\begin{eqnarray} \label{EC2}
C_1 & =&
a^Iba^J+ \sum_{i \in I'}a^iba^{L_i}(a-1)a^J +
\sum_{j \in J'}a^{I}(a-1)a^{M_j}ba^j \geq 0.
\end{eqnarray}

The pairs $(I,J)$ as above have been
completely described in \cite{KRR}.
More precisely, starting with the
chain of positive distinct divisors of $n$:
\begin{eqnarray*}
k_{0} &=& 1 \mid k_{1} \mid k_{2} \mid \ldots \mid k_{s} =n,
\end{eqnarray*}
let us consider
the two polynomials $a^I$ and $a^J$
defined by:
\begin{eqnarray} \label{EK1}
a^{I} & = & \prod_{j \; {\rm even} \;, 1 \leq j \leq s}
\frac{(a^{k_{j}}-1)}{(a^{k_{j-1}}-1)}, \quad
a^{J}= \prod_{j \; {\rm odd} \;, 1 \leq j \leq s}
\frac{(a^{k_{j}}-1)}{(a^{k_{j-1}}-1)}.
\end{eqnarray}
In \cite{KRR},
Krasner and Ranulac proved
that a pair
$(I,J)$ satisfies Eqs.(\ref{EK1})
if and only if
for any $z \in \{0, \ldots , n-1 \}$ there exists a unique
$(i,j)$, with $i \in I$ and $j \in J$,
such that
$i+j=z$, i.e., $a^Ia^J=\frac{a^{n}-1}{a-1}$. The pair
$(I,J)$ is called a
{\it Krasner factorization (of order $n$)}.

%------------------------------------------
\section{Factorizations of cyclic groups and positively factorizing codes} \label{FPF}

%-------------------------------------------
\subsection{Haj\'{o}s factorizations}
\label{KHF}

In
\cite{HA50}, Haj\'{o}s gave a method,
slightly corrected later by Sands
in \cite{SA}, for the construction of
a class
of factorizations of an abelian group $(G,+)$
which are of special interest in the construction
of factorizing codes and
in the proof of
our results.
As in
\cite{DF96}, we describe this method
for the cyclic group
$\Z_{n}$ of order $n$
(Definition
\ref{HCG}).
The
corresponding
factorizations will be named
{\it Haj\'{o}s factorizations}.

For subsets
$S= \{ s_{1}, \ldots , s_{q} \}$, $T$ of $\Z_{n}$,
we define
$S \circ T$ as the family of subsets of $\Z_{n}$
having the form
$\{ s_{i}+t_{i} \mid i \in \{ 1, \ldots , q \} \}$, where
$\{ t_{1}, \ldots ,t_{q} \}$ is any multiset of elements of $T$
having the same cardinality as $S$.

\begin{definition} \label{HCG}
Let $R,T$ be subsets of $\N$.
$(R,T)$ is a Haj\'{o}s factorization of
$\Z_{n}$ if and only if there
exists
a chain of positive distinct
divisors of $n$
\begin{eqnarray} \label{EH1}
k_{0} &=& 1 \mid k_{1} \mid k_{2} \mid \cdots \mid k_{s} =n,
\end{eqnarray}
such that
\begin{eqnarray} \label{EH2}
a^{R} & \in &
( \cdots (( \frac{a-1}{a-1} \cdot \frac{a^{k_1}-1}{a-1}) \circ
\frac{a^{k_{2}}-1}{a^{k_1}-1}) \cdot \ldots \circ \ldots
\frac{a^{n}-1}{a^{k_{s-1}}-1} ),
\end{eqnarray}
\begin{eqnarray} \label{EH3}
a^{T} & \in &
( \cdots (( \frac{a-1}{a-1} \circ \frac{a^{k_1}-1}{a-1}) \cdot
\frac{a^{k_{2}}-1}{a^{k_1}-1}) \circ \ldots \cdot \ldots
\frac{a^{n}-1}{a^{k_{s-1}}-1} ),
\end{eqnarray}
Furthermore we have
$R,T \subseteq \{0, \ldots , n-1 \}$.
\end{definition}

We now recall three results
which will be used.
We begin with a
recursive construction of
Haj\'{o}s factorizations of
$\Z_{n}$, which
was first given in
\cite{LAM97} as a direct result and later
proved in \cite{DF06} for the sake
of completeness.

\begin{proposition} \label{HR} \cite{LAM97}
Let $R,T \subseteq \{0, \ldots , n-1 \}$
and suppose that $(R,T)$
is
a Haj\'{o}s factorization of
$\Z_{n}$
with respect to the
chain
$k_{0} = 1 \mid k_{1} \mid k_{2} \mid \cdots \mid k_{s} =n$
of positive distinct divisors of $n$.
Then either $(R,T) = (R_1, T_1)$
or $(R,T) = (T_1, R_1)$,
where $(R_1, T_1)$
satisfies one of the two following
conditions.
\begin{enumerate}
\item[1)]
There exists $t \in \{0, \ldots , n-1 \}$
such that
$R_1 =\{0, \ldots , n-1 \}$ and
$T_1 =\{ t \}$.
Furthermore, $s=1$.
\item[2)]
$R_1=R^{(1)} + \{0, 1, \ldots , g-1\}h$,
$T_1=T^{(1)} \circ \{0, 1, \ldots , g-1\}h$,
$(R^{(1)}, T^{(1)})$ being a Haj\'{o}s factorization
of $\Z_{h}$, $g,h \in \N$,
$n=gh$,
$R^{(1)}, T^{(1)} \subseteq \{0, \ldots , h-1 \}$.
The chain of divisors defining
$(R^{(1)}, T^{(1)})$ is
$k_{0} = 1 \mid k_{1} \mid k_{2} \mid \cdots \mid k_{s-1}=h$.
\end{enumerate}
\end{proposition}

\begin{example} \label{ERC}
{\rm The pair $(\{0,1\}, \{1\})$
is a Haj\'{o}s factorization of
$\Z_{2}$ (condition
1) in Proposition
\ref{HR}). Thus, $(\{1,2\}, \{1,3,5\})$
is a Haj\'{o}s factorization of
$\Z_{6}$ since
$\{1,2\} = \{0,1\} \circ \{0,1,2\}2$
and $\{1,3,5\} = \{1\} + \{0,1,2\}2$
(condition 2) in Proposition
\ref{HR}). Finally,
$(\{1,2,7,8\}, \{1,3,5\})$
is a Haj\'{o}s factorization of
$\Z_{12}$ since
$\{1,2,7,8\} = \{1,2\} + \{0,1\}6$
(condition 2) in Proposition
\ref{HR}).}
\end{example}

As observed in \cite{DF96},
the simplest example of Haj\'{o}s factorizations
is given by Krasner pairs.
Thus, Proposition
\ref{HR} can also be applied:
for each $n > 1$,
for each Krasner pair $(I,J)$ of order $n$,
there exist
$h, g \in \N$, with
$h<n=gh$ such that either
$I=I^{(1)}+ \{0,1, \ldots, (g-1)\}h$,
with $(I^{(1)},J)$ being a Krasner factorization
of $\Z_{h}$
or $J=J^{(1)}+ \{0,1, \ldots, (g-1)\}h$,
with $(I, J^{(1)})$ being a Krasner factorization
of $\Z_{h}$.

\begin{example} \label{ERCK}
{\rm The pair $(\{0,1\}, \{0\})$
is a Krasner factorization of
$\Z_{2}$.
Then, $(\{0,1\}, \{0,2,4\})$
is a Krasner factorization of
$\Z_{6}$ and
$\{0,2,4\} = \{0\} + \{0,1,2\}2$
(condition 2) in Proposition
\ref{HR}). Finally,
$(\{0,1,6,7\}, \{0,2,4\})$
is a Krasner factorization of
$\Z_{12}$ and
$\{0,1,6,7\} = \{0,1\} + \{0,1\}6$
(condition 2) in Proposition
\ref{HR}).}
\end{example}

A stronger relationship
between Haj\'{o}s factorizations
and Krasner pairs
is reported in
Theorem \ref{HC} below and makes some equations
between polynomials
in $\N[a]$
intervene. This result, along with Lemma \ref{L72},
will be needed.

\begin{theorem} \label{HC} \cite{DF96}
Let $(R,T)$ be subsets of
$\{0, \ldots , n-1 \}$. The
following conditions are equivalent
\begin{itemize}
\item [1)]
$(R,T)$ is a Haj\'{o}s factorization of
$\Z_{n}$.
\item [2)]
There exists
a Krasner factorization $(I,J)$ of
$\Z_{n}$ such that $(I,T)$, $(R,J)$ are
(Haj\'{o}s) factorizations of $\Z_{n}$.
\item [3)]
There exist
$L,M \subseteq \N$ and a
Krasner factorization $(I,J)$ of
$\Z_{n}$
such that
\begin{eqnarray} \label{EF}
a^R &=& a^{I}(1+a^{M}(a-1)), \quad
a^T=a^{J}(1+a^{L}(a-1)).
\end{eqnarray}
\end{itemize}
Furthermore,
$2) \Leftrightarrow 3)$ also
holds for $R,T \subseteq \N$.
\end{theorem}

A construction of sets $L,M$ satisfying Eq.(\ref{EF}) may be found in \cite{DF89}.
Theorem \ref{HC} points out that for each
Haj\'{o}s factorization $(R,T)$, there is
a Krasner factorization $(I,J)$ associated with
$(R,T)$. In \cite{LAM97}
$(I,J)$ is called
a {\it Krasner companion
factorization} of
$(R,T)$. 
Each
Krasner companion factorization of
a given Haj\'{o}s factorization
$(R,T)$ is associated
with a chain of divisors of $n$ defining
$(R,T)$ and can be easily constructed starting with it
(see \cite[Proposition 4.2] {DF05}). 
It is worth pointing out that
if $I,J,R,T$ satisfy Eqs.(\ref{EF})
(and condition 2) in Proposition \ref{HR})
then $I,R$ (or $J,T$) are such that
$I=I^{(1)}+ \{0,1, \ldots, (g-1)\}h$,
$R=R^{(1)}+ \{0,1, \ldots, (g-1)\}h$, where
$I^{(1)}$, $R^{(1)}$, $g$, $h$ satisfy all the
other conditions reported in
Proposition \ref{HR}
(for a proof, see the more general statement in \cite[Lemma 4.5] {DF07a}). Finally,
looking at Definition \ref{HCG},
we see that for a Haj\'{o}s factorization
$(R,T)$ of
$\Z_{n}$,
we have $R,T \subseteq \{0, \ldots , n-1 \}$.
Therefore, in what follows,
for $R,T \subseteq \N$,
we will say that $(R,T)$ is a Haj\'{o}s
factorization
of
$\Z_{n}$ if
$(R_{(n)}, T_{(n)})$
satisfies the conditions in
Definition \ref{HCG} where,
for a subset $X$
of $\N$ and
$n \in \N$, we
denote
$X_{(n)}= \{x' ~|~ 0 \leq x' \leq n-1, \;
\exists x \in X, x=x' \pmod{n}\}$.
This is equivalent, as
Lemma \ref{L72} shows,
to define
Haj\'{o}s
factorizations of
$\Z_{n}$
as those pairs satisfying
Eqs.(\ref{EF}).

\begin{lemma} \cite{DF01} \label{L72}
Let $(I,J)$ be a Krasner factorization
of $\Z_{n}$.
Let $R,R',M$ be subsets
of $\N$
such that $a^R=a^I(1+a^M(a-1))$ and
$a^{R'}=a^{R_{(n)}}$.
Then,
$M' \subseteq \N$
exists
such that
$a^{R'}=a^I(1+a^{M'}(a-1))$ and
$I + max ~ M' + 1 \subseteq \{ 0, \ldots , n-1 \}$.
Furthermore, if we set $R'=\{ r_1, \ldots , r_q \}$,
$R = \{r_1+ \lambda_1 n, \ldots , r_q + \lambda_q n \}$,
for $\lambda_1, \ldots \lambda_q \geq 0$,
and if we set
$a^H= a^{r_1+\{0,n, \ldots, (\lambda_1 -1)n\}}+
\ldots + a^{r_q+\{0,n, \ldots, (\lambda_q -1)n\}}$
then we have a disjoint union $M=M' \cup M^{''}$
with
$M^{''} \subseteq \N$,
$a^{M''} =a^Ja^H$ and
$a^{R}=
a^{R'}+a^I(a-1)a^{M^{''}}$.
\end{lemma}

%--------------------------------------------------
\subsection{Good arrangements} \label{GAS}

We now give a brief exposition of results
which relate factorizing codes
and factorizations of cyclic groups,
through the notion of a {\it good arrangement}.
We follow the notations used in
\cite{DF05} where matrices with entries in $A^*$ or
in $\N$ will be
considered.
Given a matrix
$\mathcal{A} = (a_{p,q})_{1 \leq p \leq m, \; 1 \leq q \leq \ell}$
with entries
in $\N$
and an integer $n$, $n \geq 2$,
we denote
$\mathcal{A}_{(n)} = (a'_{p,q})_{1 \leq p \leq m, \; 1 \leq q \leq \ell}$,
where, for each $p, q$,
$1 \leq p \leq m, \; 1 \leq q \leq \ell$,
we have $a'_{p,q} = a_{p,q} \pmod{n}$,
$0 \leq a'_{p,q} \leq n-1$.
We also denote
$h + \mathcal{A} = (b_{p,q})_{1 \leq p \leq m, \; 1 \leq q \leq \ell}$,
where, for each $p, q$,
$1 \leq p \leq m, \; 1 \leq q \leq \ell$,
we have $b_{p,q} = h + a_{p,q}$
and $\mathcal{A} \cup \mathcal{B} =
(a_{p,q})_{1 \leq p \leq m, \; 1 \leq q \leq 2\ell}$,
where $\mathcal{B} =
(a_{p,q})_{1 \leq p \leq m, \; \ell+1 \leq q \leq 2\ell}$.
Finally $\cup_{i=1}^n \mathcal{A}_i = (\cup_{i=1}^{n-1} \mathcal{A}_i) \cup
\mathcal{A}_n$. A dual operation of union with respect to
the columns is assumed to be defined.

An arrangement of $X$, with $X \subseteq A^*$
(resp. $X \subseteq N$),
will be an
arrangement of the elements of $X$ in a matrix
with entries in $A^*$ (resp. $\N$) and
size $\Card(X)$.

\begin{definition} \cite{DF05} \label{GA}
Let $(R_1,T_1), \ldots , (R_m,T_m)$
be Haj\'{o}s
factorizations
of
$\Z_{n}$ having $(I,J)$
as a Krasner companion factorization.
An arrangement $\mathcal{D} = (r_{p,q})_{1 \leq p \leq m, \; 1 \leq q \leq
l}$
of $\cup_{p=1}^m R_p$
having the $R_p$'s as rows
is a {\rm good arrangement} of
$(R_1, \ldots , R_m)$ ({\rm with respect to the rows})
if $\mathcal{D}$ can be recursively constructed
by using the following three rules.
\begin{enumerate}
\item[1)]
$\mathcal{D}$ is a good arrangement of
$\cup_{p=1}^m R_p$
if
$\mathcal{D}_{(n)}$ is a good arrangement of
$\cup_{p=1}^m (R_p)_{(n)}$.
\item[2)]
Suppose that $(R_p,T_p)$ satisfies condition $1)$ in
Proposition \ref{HR}, for all $p \in \{1, \ldots , m \}$.
If $R_p = \{ r_p \}$ with $r_p \in \{0, \ldots , n-1 \}$,
then $\mathcal{D}$
is the matrix with only one column
having $r_p$ as the $p$th entry.
If $R_p  =\{r_{p,0}, \ldots , r_{p,n-1} \}$
with $r_{p,i} = i$, then
$\mathcal{D} =
(r_{p,j})_{1 \leq p \leq m, \; 0 \leq j \leq n-1}$.
\item[3)]
Suppose that $(R_p,T_p)$ satisfies
condition $2)$ in
Proposition \ref{HR}, for all $p \in \{1, \ldots , m \}$, i.e.,
either
$R_p=R^{(1)}_{p} + \{0, h, \ldots , (g-1)h)\}$
or
$R_p=R^{(1)}_{p} \circ \{0, h, \ldots , (g-1)h)\}$.
Let $\mathcal{D}^{(1)}$ be a good arrangement of
$\cup_{p=1}^m R^{(1)}_p$ having the $R^{(1)}_p$'s as rows.
In the first case, we set
$\mathcal{D} = \cup_{k = 0}^{g-1} (kh + \mathcal{D}^{(1)})$.
In the second case,
$\mathcal{D}$
is obtained by taking $\mathcal{D}^{(1)}$
and then substituting in it each $r_{p,q}^{(1)} \in R_p^{(1)}$
with the corresponding $r_{p,q}^{(1)}+ \lambda_{p,q} h \in R_p$.
\end{enumerate}
\end{definition}

Let $(R_1,T_1), \ldots , (R_m,T_m)$
be Haj\'{o}s
factorizations
of
$\Z_n$ having $(I,J)$
as a Krasner companion factorization.
Obviously, we can
consider arrangements of
$\cup_{p=1}^m R_p$
having the $R_p$'s as {\it columns} and
therefore,
we can
give a dual
notion of a {\it good arrangement} of
$\cup_{p=1}^m R_p$
{\it with respect to the columns}
(by using the corresponding dual operation
$\cup$).
This arrangement
will be
the transpose matrix of a good
arrangement of
$\cup_{p=1}^m R_p$ with respect to the rows.
In \cite{DF05} the author proved that there exists
a {\it unique} good arrangement of $\cup_{p=1}^m R_p$
with respect to the rows (resp. columns).
In the same paper \cite{DF05}, the following property of good arrangements
has been proved.

\begin{proposition} \label{GAP}
Let $(R_1,T_1), \ldots , (R_m,T_m)$
be Haj\'{o}s
factorizations
of
$\Z_{n}$ having $(I,J)$
as a Krasner companion factorization.
Let $\mathcal{D}
= (r_{p,q})_{1 \leq p \leq m, \; 1 \leq q \leq \ell}$
be the good arrangement of $\cup_{p=1}^m R_p$
with respect to the rows. Then, the two
following conditions are satisfied.
\begin{itemize}
\item[a)]
For each column $W_q=(r_{1,q}, \ldots , r_{m,q})$
of $\mathcal{D}$, there is
an ordered sequence $J_q=(j_{1,q}, \ldots , j_{m,q})$
of elements of $J$ satisfying:
\begin{eqnarray} \label{EQG1}
r_{1,q}+j_{1,q} &=& r_{2,q} + j_{2,q} = \ldots =
r_{m,q}+  j_{m,q} = n_q \pmod{n}.
\end{eqnarray}
\item[b)]
Suppose that $R_p,T_q \subseteq \{0, \ldots , n-1 \}$.
Thus, for each column $W_q=(r_{1,q}, \ldots , r_{m,q})$
of $\mathcal{D}$, there exists
an ordered sequence $J_q=(j_{1,q}, \ldots , j_{m,q})$
of elements of $J$ satisfying:
\begin{eqnarray} \label{EQG2}
r_{1,q}+j_{1,q} &=& r_{2,q} + j_{2,q} = \ldots =
r_{m,q}+  j_{m,q} = n_q.
\end{eqnarray}
\end{itemize}
The $n_q$'s are all different.
\end{proposition}

\begin{definition} \cite{DF05} \label{IA}
Let
$C_1 = (a^{r_{p,q}}ba^{v_{p,q}})_{1 \leq p \leq m, \; 1 \leq q \leq \ell}$
be an arrangement
of $C_1 \subseteq a^*ba^*$.
The matrix $\mathcal{R} =
(r_{p,q})_{1 \leq p \leq m, \; 1 \leq q \leq \ell}$
is the {\rm induced arrangement} of the rows
$R_{p} =\{r_{p,q} ~|~ q \in \{1, \ldots , \ell \} \}$
and the matrix $\mathcal{T} =
(v_{p,q})_{1 \leq p \leq m, \; 1 \leq q \leq \ell}$
is the {\rm induced arrangement} of the columns
$T_{q} =\{v_{p,q} ~|~ p \in \{1, \ldots , m \} \}$.
Furthermore, $R_{p,w} =
\{a^{r_{p,q}}ba^{v_{p,q}} ~|~ 1 \leq q \leq \ell \}$
(resp. $T_{q,w} =
\{a^{r_{p,q}}ba^{v_{p,q}} ~|~ 1 \leq p \leq m \}$)
is
a {\rm word-row} (resp.
a {\rm word-column})
of $C_1$, for $1 \leq p \leq m$
(resp. $1 \leq q \leq \ell$).
\end{definition}

\begin{definition} \cite{DF05} \label{GAC}
An arrangement
$C_1 =
(a^{r_{p,q}}ba^{v_{p,q}})_{1 \leq p \leq m, \; 1 \leq q \leq \ell}$
of $C_1 \subseteq a^*ba^*$
is a {\rm good arrangement
(with $(I,J)$ as
a Krasner associated
pair)}
if it satisfies the
following three conditions:
\begin{enumerate}
\item[1)]
For each row $R_p$ and each column
$T_q$, $1 \leq p \leq m$, $1 \leq q \leq \ell$,
$(R_p,T_q)$ is a Haj\'{o}s factorization of
$\Z_{n}$ having $(I,J)$ as a Krasner
companion factorization
with respect to a chain of divisors
of $n = \Card(C_1)$.
\item[2)]
The induced arrangement of the rows
is a good arrangement
of $\cup_{p=1}^m R_p$
with respect to the rows.
\item[3)]
The induced arrangement of the columns
is a good arrangement
of $\cup_{q=1}^{\ell} T_q$
with respect to the columns.
\end{enumerate}
We set $\mathcal{G}_1(I,J) = \{C_1 \subseteq a^*ba^* ~|~ $
there exists a good
arrangement of $C_1$
with $(I,J)$ as a Krasner associated
pair$\}$.
\end{definition}

To simplify notation, from now on we will write
$\mathcal{G}(I,J)$ instead of $\mathcal{G}_1(I,J)$.
The remainder of this section will be devoted to
a list of results on good arrangements and
factorizing codes.

\begin{proposition} \cite{DF05} \label{arrangement}
Let $C_1$ be a subset of $a^*ba^*$
which satisfies Eq.(\ref{EC2})
with $(I,J)$ being a
Krasner factorization
of $\Z_{n}$.
Then,
there exists a good
arrangement of $C_1$
with $(I,J)$ as a Krasner associated
pair.
\end{proposition}

\begin{remark}
{\rm Observe that, in view of Proposition \ref{arrangement},
if $C_1$ satisfies Eq.(\ref{EC2})
then $C_1$ is a polynomial with coefficients $0,1$,
i.e., $C_1 \subseteq a^*ba^*$.}
\end{remark}

If $C_1 = C \cap a^*ba^*$ for a factorizing code $C$, then
$C_1$ has a good arrangement. Proposition below shows a particular case
in which the converse holds. In general this is not true. Indeed,
there exist
sets $C_1$ having a good arrangement and
which are not codes (see \cite{DF07b}).

\begin{proposition} \label{main1} \cite{DF07b}
Let $1 ~ | ~ h ~ | ~ hg =n$ be a chain
of divisors of $n$, let $(I,J)$ be a Krasner
pair of order $n$, with
$I = \{0, 1, \ldots , h-1 \}$,
$J = \{0, h, \ldots , (g-1)h \}$
and let $C_1 \subseteq a^*ba^*$.
Assume that $C_1$ has a good arrangement with
$(I,J)$ as a Krasner associated pair.
Then,
there exist
$t_i \in \{0, 1, \ldots , h -1 \}$,
$\lambda_{i,k} \in \{0, 1, \ldots , g -1 \}$,
$0 \leq i \leq h - 1$,
$0 \leq k \leq g - 1$,
such that
\begin{eqnarray} \label{E1}
C_1^{\! \! \! \pmod{n}} & = & \sum_{k = 0}^{g - 1} \sum_{i = 0}^{h - 1}
a^{i + \lambda_{i,k}h} b a^{t_i + kh}.
\end{eqnarray}
Furthermore,
there exist factorizing codes
$C$, $C'$ such that
$C \cap a^*ba^* = C_1^{\! \! \! \pmod{n}}$,
$C' \cap a^*ba^* = C_1$,
and $a^n \in C \cap C'$.
\end{proposition}

In the statement below $\Omega(n)$ will denote the
{\it number of factors} in the prime factorization
of $n$ \cite{HW}.

\begin{corollary} \label{main2} \cite{DF07b}
Let $C_1$ be a subset of $a^*ba^*$,
with $\Card(C_1) = n$ and
$\Omega(n) \leq 2$.
Then,
there exists
a factorizing code $C$ such that
$C_1 = C \cap a^*ba^*$ if and only if
there exists a Krasner factorization
$(I,J)$ of
$\Z_n$
such that
$C_1 \in \mathcal{G}(I,J)$.
\end{corollary}

We recall that under the hypothesis of Proposition \ref{main1},
the class $\mathcal{G}(I,J)$ coincide with other classes of subsets
di $a^*ba^*$ defined in \cite{DF07a}.

%-------------------------------------
\section{Factorizations of cyclic groups and finite maximal codes} \label{FMC}

As pointed out in \cite{BPR}, for any finite maximal code $X$ one can associate
with each letter $a$ several factorizations of $\Z_n$, where $n$ is the order of $a$.
A word $w$ is {\it right completable}
in $X^*$ if there exists $v \in A^*$ such that $wv \in X^*$. 
The following is Theorem 12.2.6 in \cite{BPR}.

\begin{theorem}
Let $X$ be a finite maximal code. Let $\phi: A^* \rightarrow M$ be the morphism
from $A^*$ onto the syntactic monoid of $X^*$ and let $K$ be the minimal ideal of $M$.
Let $a$ be a letter and let $n$ be its order. For $u,v \in A^*$ let
$$R(u) = \{i \geq 0 ~|~ ua^i \in X^* \}, \quad L(v) = \{ j \geq 0 ~|~ a^jvA^* \cap X^* \not = \emptyset \},$$
and let $\overline{R}(u)$, $\overline{L}(v)$ denote the sets of residues $\pmod{n}$
of $R(u)$, $L(v)$. If $u,v \in \phi^{-1}(K)$ and $u$ is right completable in $X^*$, then
$(\overline{R}(u), \overline{L}(v))$ is a factorization of $\Z_n$. Moreover
$\Card(\overline{L}(v)))$ is a multiple of the degree of $X$.
\end{theorem}

This result has been enhanced in \cite{ZHSH} through the notions of left and right sets,
introduced in the same paper and recalled below.
Set $T = \{0, \ldots , n-1 \}$ and $|X| = \max\{|x| ~|~ x \in X \}$. A word $w$ is {\it strongly right completable}
for $X$
if, for all $u \in A^*$,  there exists $v \in A^*$ such that $wuv \in X^*$.

\begin{definition} \cite{ZHSH}
The set $a^P$ is a {\rm left set} of $X$ if there is a strongly right completable word
$y \in A^*$ for $X$ such that 
$$P = \{ i \in T ~|~ ya^{2n|X| + i} \in X^* \}.$$
The set $a^Q$ is a {\rm right set} of $X$ if there is a word $x \in A^*$ such that
\begin{eqnarray*}
Q &=& \{ k \in T ~|~ a^{k +2n|X|}xA^* \cap X^* \not = \emptyset \} \quad \mbox{and} \\
\forall i,j \in Q, i < j && a^{j-i} \not \in (X^*)^{-1}X^*
\end{eqnarray*}
The word $y$ (resp. $x$) is the generator of the left (resp. right) set $a^P$ (resp. $a^Q$).
\end{definition}

\begin{theorem} \label{ZH1} \cite{ZHSH}
Let $X \subseteq A^*$ be a finite maximal code and let $n$ be the order
of $a \in A$. For any left set $a^P$ of $X$ and any right set $a^Q$ of $X$,
the pair $(P,Q)$ is a factorization of $\Z_n$.
\end{theorem}

Let $a$ be a letter in $A$. In the following we set $B = A \setminus \{ a \}$
and, for $w \in B(a^*B)^*$,
$$X_w = (a^*wa^* \cap X^*) \setminus [a^n(a^*wa^* \cap X^*) \cap (a^*wa^* \cap X^*) a^n].$$
The following is part of one of the main results in \cite{ZHSH}.

\begin{theorem} \label{ZHmain} \cite{ZHSH}
Let $X \subseteq A^*$ be a finite maximal code and let $n$ be the order
of $a \in A$. Let $a^P = \{a^{p_1}, \ldots , a^{p_s} \} $ be a left set of $X$ and let
$a^Q = \{a^{q_1}, \ldots , a^{q_t} \} $ be a right set of $X$.
For any $w \in B(a^*B)^*$,
there exists an arrangement
$X_w = (a^{i_{k,m}}wa^{j_{k,m}})_{1 \leq k \leq s, \; 1 \leq m \leq t}$
of $X_w$ satisfying the following properties
\begin{itemize}
\item[(1)]
there exists
an ordered sequence $P_m=(p_{1,m}, \ldots , p_{s,m})$
of elements of $P$, $\quad 1 \leq m \leq t$, satisfying:
\begin{eqnarray} \label{EQZH1}
i_{1,m}+p_{1,m} &=& i_{2,m} + p_{2,m} = \ldots =
i_{s,m}+  p_{s,m} = q_m \pmod{n}.
\end{eqnarray}
\item[(2)]
there exists
an ordered sequence $Q_k=(q_{k,1}, \ldots , q_{k,t})$
of elements of $Q$, $\quad 1 \leq k \leq s$, satisfying:
\begin{eqnarray} \label{EQZH2}
j_{k,1} + q_{k,1} &=& j_{k,2} + q_{k,2} = \ldots =
j_{k,t} + q_{k,t} = p_k \pmod{n}.
\end{eqnarray}
\item[(3)]
For each row $R_k = \{ i_{k,m} ~|~ 1 \leq m \leq t \}$
and each column $T_m = \{j_{k,m}  ~|~ 1 \leq k \leq s \}$,
$a^{\overline{R_k}}$ is a right set and
$a^{\overline{T_m}}$ is a left set, and
$(R_k, T_m), (R_k, P), (Q, T_m)$ are factorizations of $\Z_n$.
\end{itemize}
\end{theorem}

For our aims we need a statement which is
an intermediate step in the proof of the above theorem.
It is reported below.

\begin{proposition} \label{ZHmainpart} \cite{ZHSH}
Let $X \subseteq A^*$ be a finite maximal code and let $n$ be the order
of $a \in A$. For any left set $a^P = \{a^{p_1}, \ldots , a^{p_s} \} $ of $X$
and any right set $a^Q = \{a^{q_1}, \ldots , a^{q_t} \} $ of $X$,
there is a bijection $\phi : P \times Q \rightarrow X_w$ defined by
$\phi((p_k, q_m)) = a^{i_{k,m}}wa^{j_{k,m}}$, where $i_{k,m}, j_{k,m}$ satisfy
Eqs.(\ref{EQZH1}), (\ref{EQZH2}), $1 \leq k \leq s$,  $1 \leq m \leq t$.
\end{proposition}

The following definition is a slight modification of a notion introduced
in \cite{ZHSH}.

\begin{definition} \cite{ZHSH}
Let $X$ be a finite maximal code, let $a$ be a letter of order $n$.
The {\rm system of factorizations} of $\Z_n$
{\rm induced by} $X$ is the set $\mathbb{P} \times \mathbb{Q}$, where
$$\mathbb{P} = \{ P \subseteq T ~|~ a^P \mbox{is a left set of } X \},
\quad \mathbb{Q} = \{ Q \subseteq T ~|~ a^Q \mbox{ is a right set of } X \}$$
\end{definition}

By Theorem \ref{ZH1}, for any $P \in \mathbb{P}$, $Q \in \mathbb{Q}$,
the pair $(P,Q)$ is a factorization of $\Z_n$.
Consider the following inequalities
\begin{eqnarray} \label{EQZH3}
\forall K \geq 0 & & \Card(\{a^iwa^j \in X_w ~|~ i + j \leq K \} \leq K + 1
\end{eqnarray}
The following is another main result in \cite{ZHSH}.

\begin{theorem} \label{ZH2} \cite{ZHSH}
Let $X$ be a finite maximal code, let $a$ be a letter of order $n$.
Let $\mathbb{P} \times \mathbb{Q}$ be the system of factorizations of $\Z_n$
induced by $X$.
If a Krasner factorization (of order $n$) $(I,J)$ is in $\mathbb{P} \times \mathbb{Q}$, 
then Eqs.(\ref{EQZH3}) hold.
\end{theorem}

The key argument in the proof of the above theorem is the existence
of an injection $\Phi : X_w \rightarrow a^I w a^J$ such that $\Phi(a^iwa^j) = a^{i'}wa^{j'}$,
with $i' \leq i, j' \leq j$.
Proposition \ref{ZH3}, needed for our aims, shows how, starting with $\Phi$, the authors
conclude the proof of Theorem \ref{ZH2}. Its proof is reported for the sake of completeness.

\begin{proposition} \label{ZH3} \cite{ZHSH}
Let $X$ be a finite maximal code, let $a$ be a letter in $A$ of order $n$.
Let $(I,J)$ be a
Krasner factorization of order $n$.
If there exists an injection $\Phi : X_w \rightarrow a^I w a^J$ such that $\Phi(a^iwa^j) = a^{m_i}wa^{\ell_j}$,
with $m_i \leq i, \ell_j \leq j$, then
Eqs.(\ref{EQZH3}) hold.
\end{proposition}
\begdim
Let $Y = \{(i,j) ~|~ a^iwa^j \in X_w \}$ and let $\phi : Y \rightarrow I \times J$ be
such that $\phi((i,j)) = (m_i, \ell_j)$, where $(m_i, \ell_j)$ is defined by $\Phi(a^iwa^j) = a^{m_i}wa^{\ell_j}$.
Clearly, Eqs.(\ref{EQZH3}) hold for $a^I w a^J$, that is
$$\Card(\{ a^{i'}wa^{j'} \in a^I w a^J ~|~ i' + j' \leq K \}) \leq K + 1.$$
Moreover, since $m_i \leq i, \ell_j \leq j$, we have
\begin{eqnarray*}
\{ a^{m_i}wa^{\ell_j} \in \Phi(X_w) ~|~ i + j \leq K \} & \subseteq &
\{ a^{m_i}wa^{\ell_j} \in \Phi(X_w) ~|~ m_i + \ell_j \leq K \} \\
& \subseteq & \{ a^{i'}wa^{j'} \in a^I w a^J ~|~ i' + j' \leq K \},
\end{eqnarray*}
which implies
\begin{eqnarray*}
\Card(\{ a^i w a^j \in X_w ~|~ i + j \leq K \}) & = &
\Card(\{ a^{m_i}wa^{\ell_j} \in \Phi(X_w) ~|~ i + j \leq K \} \\
& \leq &
\Card(\{ a^{i'}wa^{j'} \in a^I w a^J ~|~ i' + j' \leq K \}) \\
& \leq & K + 1
\end{eqnarray*}

The following two corollaries of Theorem \ref{ZH2} have been stated in \cite{ZHSH}.

\begin{corollary} \label{ZHCor1}
If $X$ is a finite maximal code and $a^p \in X$, where
$p$ is a prime number, then Eqs.(\ref{EQZH3}) hold.
\end{corollary}

\begin{corollary} \label{ZHCor2}
If $X$ is a finite maximal code, $(P,Q)$ is in its system of factorizations, and one among
$P,Q$ is a singleton, then Eqs.(\ref{EQZH3}) hold.
\end{corollary}

%----------------
\section{Finite maximal codes and good arrangements} \label{MCGA}

Results in this section show that Theorem \ref{ZH2} may be proved under
a weaker hypothesis. This is due to the close relation
between Eqs.(\ref{EQZH1}),(\ref{EQZH2}) and
Eqs.(\ref{EQG1}).
In Section \ref{GAS} we introduced good arrangements for subsets $C_1 \subseteq a^*ba^*$.
Now we refer to an extension of this notion to subsets $X_w$ associated with a finite maximal code
$X$. Given an arrangement of $X_w$, we consider the matrix obtained by changing $w$ with $b$ in
all its words. The former arrangement is good if so is the latter.

Let $X$ be a finite maximal code. In this section we prove the following:
\begin{itemize}
\item[(1)]
If the hypothesis of Theorem \ref{ZH2} is verified, i.e., a Krasner factorization $(I,J)$
is in the system of factorizations of $\Z_n$
induced by $X$, then there exists a good arrangement of $X_w$, for any $w \in B(a^*B)^*$,
with $(I,J)$ as a Krasner associated pair
(Theorem \ref{ZH2DF}).
\item[(2)]
For any $w \in B(a^*B)^*$, if there exists a good arrangement of $X_w$ with
$(I,J)$ as a Krasner associated pair, then the hypothesis of Proposition \ref{ZH3} is
verified, i.e., there exists an injection $\Phi : X_w \rightarrow a^I w a^J$ such that $\Phi(a^iwa^j) = a^{m_i}wa^{\ell_j}$,
with $m_i \leq i, \ell_j \leq j$ (Proposition \ref{ZH3DF}).
\item[(3)]
If the hypothesis of Corollary \ref{ZHCor1} is satisfied, i.e., $a^p \in X$, where
$p$ is a prime number, then there exists a good arrangement of $X_w$, for any $w \in B(a^*B)^*$
(Corollary \ref{ZHDFCor1}).
\item[(4)]
If the hypothesis of Corollary \ref{ZHCor2} is satisfied, i.e., $(P,Q)$
is in the system of factorizations of $\Z_n$
induced by $X$ and one among
$P,Q$ is a singleton, then there exists a good arrangement of $X_w$, for any $w \in B(a^*B)^*$
(Corollary \ref{ZHDFCor2}).
\end{itemize}

Recall that $\overline{i}$ denotes the unique integer in $\{0, \ldots, n -1 \}$
such that $i = \overline{i} \pmod{n}$.
The following result is needed for the proof of Theorem
\ref{ZH2DF}.

\begin{proposition} \label{propZHDF}
If $X$ is a set of words such that
\begin{itemize}
\item[(1)]
there is an arrangement
$X_w = (a^{i_{k,m}}wa^{j_{k,m}})_{1 \leq k \leq s, \; 1 \leq m \leq t}$
of $X_w$ such that for each row $R_k = \{ i_{k,m} ~|~ 1 \leq m \leq t \}$
and each column $T_m = \{j_{k,m}  ~|~ 1 \leq k \leq s \}$,
the pairs
$(R_k, T_m), (R_k, J), (I, T_m)$ are factorizations of $\Z_n$, where
$(I,J)$ is a Krasner factorization of $\Z_n$ .
\item[(2)]
there exists
a bijection $\phi : J \times I \rightarrow X_w$ such that
$\phi(m,k) = a^{i_{k,m}}wa^{j_{k,m}}$,
\end{itemize}
then there exists a good arrangement of $X_w$.
\end{proposition}
\begdim
By Definition \ref{GAC}, an arrangement  of $X_w = (a^{i_{k,m}}wa^{j_{k,m}})_{1 \leq k \leq s, \; 1 \leq m \leq t}$
of $X_w$ is good if the induced arrangement
$\overline{X}_w = (a^{\overline{i_{k,m}}}wa^{\overline{j_{k,m}}})_{1 \leq k \leq s, \; 1 \leq m \leq t}$
is good.
Hence, we may assume $i_{k,m}, j_{k,m} \in \{0, \ldots , n-1 \}$.
The proof is by induction on the length $s$ of the chain
of positive distinct divisors of $n$ associated with the rows $R_k$, the columns
$T_m$, $I$ and $J$.

If $s = 1$, then $X_w = (a^{i_{1,m}}wa^{j_{1,m}})_{1 \leq m \leq n}$
with $i_{1,m} = m -1  \pmod{n}$, $j_{1,m} \in \{0, \ldots , n-1 \} = I$, and
$J = \{0 \}$ or $X_w =(a^{i_{k,1}}wa^{j_{k,1}})_{1 \leq k \leq n}$
with $j_{k,1} = k -1  \pmod{n}$, $i_{k,1} \in \{0, \ldots , n-1 \} = J$,
and $I = \{0 \}$. In both cases, by Definition \ref{GAC}, this is a good arrangement of $X_w$.

Assume $s > 1$.
We may assume $I=I^{(1)}+ \{0,1, \ldots, (g-1)\}h$,
$R_k=R_k^{(1)} + \{0, 1, \ldots , g-1\}h$,
$T_m=T_m^{(1)} \circ \{0, 1, \ldots , g-1\}h$, where $n = gh$,
$(R_k^{(1)}, T_m^{(1)})$ are Haj\'{o}s factorizations
of $\Z_{h}$ having $(I^{(1)}, J)$ as a Krasner companion factorization
(a similar argument applies in the other cases). Moreover,
$R_k^{(1)}, T_m^{(1)} \subseteq \{0, \ldots , h-1 \}$ and
the chain of divisors defining
$(R_k^{(1)}, T_m^{(1)})$ has length $s-1$.
Consider the restriction $\phi_t$ of $\phi$ to $J \times I^{(1)} +th$, $t \in \{0, 1, \ldots , g-1\}$
and the corresponding submatrices $X^{(t)}_w$ of $X_w$.
By induction hypothesis, there is a good arrangement
of $X^{(t)}_w$.
Thus $\cup _{t = 0}^{g-1} X^{(t)}_w$
is the required good arrangement of $X_w$.
\enddim

\begin{theorem} \label{ZH2DF}
Let $X$ be a finite maximal code.
If a Krasner factorization $(I,J)$ of $\Z_n$ is
is in the system of factorizations of $\Z_n$
induced by $X$, then there exists a good arrangement of $X_w$, for any $w \in B(a^*B)^*$,
with $(I,J)$ as a Krasner associated pair.
\end{theorem}
\begdim
Let $X$ be a finite maximal code. Assume that a Krasner factorization $(I,J)$ of $\Z_n$ 
is in the system of factorizations of $\Z_n$
induced by $X$. By Theorem \ref{ZHmain} and Proposition \ref{ZHmainpart}, there exists
a bijection $\phi : J \times I \rightarrow X_w$ which induces an arrangement
$X_w = (a^{i_{k,m}}wa^{j_{k,m}})_{1 \leq k \leq s, \; 1 \leq m \leq t}$
of $X_w$ such that for each row $R_k = \{ i_{k,m} ~|~ 1 \leq m \leq t \}$
and each column $T_m = \{j_{k,m}  ~|~ 1 \leq k \leq s \}$,
the pairs
$(R_k, T_m), (R_k, J), (I, T_m)$ are factorizations of $\Z_n$.
These are all Haj\'{o}s factorizations of
$\Z_{n}$ having $(I,J)$ as a Krasner
companion factorization (Theorem \ref{HC}).
Therefore, by Proposition \ref{propZHDF}, the conclusion follows.
\enddim

\begin{proposition} \label{ZH3DF}
Let $w \in B(a^*B)^*$.
If there exists a good arrangement of $X_w$, with
$(I,J)$ as a Krasner associated pair, then
there exists an injection $\Phi : X_w \rightarrow a^I w a^J$ such that $\Phi(a^iwa^j) = a^{\lambda_i}wa^{\sigma_j}$,
with $\lambda_i \leq i, \sigma_j \leq j$. Consequently,
Eqs.(\ref{EQZH3}) hold.
\end{proposition}
\begdim
Let $w \in B(a^*B)^*$.
Assume that there exists a good arrangement
$X_w = (a^{r_{p,q}}ba^{v_{p,q}})_{1 \leq p \leq m, \; 1 \leq q \leq \ell}$ of $X_w$,  with
$(I,J)$ as a Krasner associated pair. Therefore, $X_w$ satisfies the three conditions in Definition
\ref{GAC}. Since there is a bijection between $X_w$
and $(a^{\overline{r_{p,q}}}wa^{\overline{v_{p,q}}})_{1 \leq p \leq m, \; 1 \leq q \leq \ell}$,
we may assume $r_{p,q}, v_{p,q} \in \{0, \ldots, n -1 \}$.

We prove the statement by induction on the length $s$ of the chain
$k_{0} = 1 \mid k_{1} \mid k_{2} \mid \cdots \mid k_{s} =n$
of positive distinct divisors of $n$ associated with the rows $R_p$, the columns
$T_q$, $1 \leq p \leq m$, $1 \leq q \leq \ell$, $I$ and $J$.
If $s = 1$, then $X_w = (a^{r_{1,q}}wa^{v_{1,q}})_{1 \leq q \leq n}$
with $r_{1,q} = q -1  \pmod{n}$, $v_{1,q} \in \{0, \ldots , n-1 \} = I$, and
$J = \{0 \}$ or $X_w =(a^{r_{p,1}}wa^{v_{p,1}})_{1 \leq p \leq n}$
with $v_{p,1} = p -1  \pmod{n}$, $r_{p,1} \in \{0, \ldots , n-1 \} = J$,
and $I = \{0 \}$. In the first case the map defined by $\Phi(a^{r_{1,q}}wa^{v_{1,q}}) =
a^{r_{1,q}}w$ is the required injection. The argument is similar in the second case.

Assume $s > 1$ and set $n = gh$, where $k_{s-1}=h$.
We may assume $I=I^{(1)}+ \{0,1, \ldots, (g-1)\}h$,
$R_p=R_p^{(1)} + \{0, 1, \ldots , g-1\}h$,
$T_q=T_q^{(1)} \circ \{0, 1, \ldots , g-1\}h$, where
$(R_p^{(1)}, T_q^{(1)})$ are Haj\'{o}s factorizations
of $\Z_{h}$ having $(I^{(1)}, J)$ as a Krasner companion factorization
(a similar argument applies in the other cases). Moreover,
$R_p^{(1)}, T_q^{(1)} \subseteq \{0, \ldots , h-1 \}$ and
the chain of divisors $k_{0} = 1 \mid k_{1} \mid k_{2} \mid \cdots \mid k_{s-1}=h$
defining
$(R_p^{(1)}, T_q^{(1)})$ has length $s-1$.
Finally, the matrices $X_w^{(t)}$ obtained by considering
words $a^{i}wa^{j}$ with $i \in R_p^{(1)}$, $t \in \{0,1, \ldots, (g-1)\}$,
$j \in T_q^{(1)}$ and $a^{i +th}wa^{j +\mu h} \in X_w$ are
good arrangements with $(I^{(1)},J)$ as a Krasner associated pair.
By induction hypothesis, there are $g$ injective functions
$\Phi_t : X_w^{(t)} \rightarrow a^{I^{(1)}} w a^J$ such that $\Phi(a^iwa^j) = a^{\lambda_i}wa^{\sigma_j}$,
with $\lambda_i \leq i, \sigma_j \leq j$.
The function $\Phi$, defined by $\Phi(a^{i +th}wa^{j +\mu h}) = a^{th} \Phi_t(a^{i}wa^{j})$
is the required function.
\enddim

\begin{corollary} \label{ZHDFCor1}
If $X$ is a finite maximal code and $a^p \in X$, where
$p$ is a prime number, then there exists a good arrangement of $X_w$, for any $w \in B(a^*B)^*$.
Consequently, Eqs.(\ref{EQZH3}) hold.
\end{corollary}
\begdim
Let $X$ be as in the statement. By Theorem \ref{ZHmain},
for any $w \in B(a^*B)^*$
there exists an arrangement
$X_w = (a^{i_{k,m}}wa^{j_{k,m}})_{1 \leq k \leq s, \; 1 \leq m \leq t}$
of $X_w$ such that $(R_k, T_m)$ is a factorization of $\Z_p$,
for each row $R_k = \{ i_{k,m} ~|~ 1 \leq m \leq t \}$
and each column $T_m = \{j_{k,m}  ~|~ 1 \leq k \leq s \}$.
Therefore, $X_w = (a^{i_{1,m}}wa^{j_{1,m}})_{1 \leq m \leq p}$
with $R_1 = \{0, \ldots , p-1 \} \pmod{p}$ or $X_w =(a^{i_{k,1}}wa^{j_{k,1}})_{1 \leq k \leq p}$
with $T_1 = \{0, \ldots , p-1 \} \pmod{p}$. By Definition \ref{GAC},
in both cases this is a good arrangement of $X_w$.
\enddim

\begin{corollary} \label{ZHDFCor2}
If $X$ is a finite maximal code, $(P,Q)$ is in its system of factorizations, and one among
$P,Q$ is a singleton, then there exists a good arrangement of $X_w$, for any $w \in B(a^*B)^*$.
Consequently, Eqs.(\ref{EQZH3}) hold.
\end{corollary}
\begdim
Let $X$ be as in the statement. By Theorem \ref{ZHmain},
for any $w \in B(a^*B)^*$
there exists an arrangement
$X_w = (a^{i_{k,m}}wa^{j_{k,m}})_{1 \leq k \leq s, \; 1 \leq m \leq t}$
of $X_w$ such that $(R_k, T_m)$ is a factorization of $\Z_p$,
for each row $R_k = \{ i_{k,m} ~|~ 1 \leq m \leq t \}$
and each column $T_m = \{j_{k,m}  ~|~ 1 \leq k \leq s \}$.
By hypothesis $s =1$, $m = n$ or $s = n$, $m =1$.
Correspondingly, $X_w = (a^{i_{1,m}}wa^{j_{1,m}})_{1 \leq m \leq n}$
with $R_1 = \{0, \ldots , n-1 \} \pmod{n}$ or $X_w =(a^{i_{k,1}}wa^{j_{k,1}})_{1 \leq k \leq n}$
with $T_1 = \{0, \ldots , n-1 \} \pmod{n}$. By Definition \ref{GAC},
in both cases this is a good arrangement of $X_w$.
\enddim

%----------------------------------------
\section{Open problems} \label{OP}

Let $X$ be a finite maximal code.

Of course, the main open problem is whether there
always exists a good arrangement of $X_w$, for any $w \in B(a^*B)^*$.
Or, following \cite{ZHSH}, if there always exists
a Krasner factorization $(I,J)$ such that $a^I$ is a left set and $a^J$ is
a right set (equivalently, $(I,J)$
is in the system of factorizations of $\Z_n$
induced by $X$).
This could be related to a recursive construction of the family of finite
maximal codes.

Other open problems are the following:

Is the converse of Theorem \ref{ZH3DF} true?
That is, does the existence of a good arrangement of $X_w$, for any $w \in B(a^*B)^*$, with
$(I,J)$ as a Krasner associated pair, imply that $(I,J)$
in the system of factorizations of $\Z_n$
induced by $X$?

Is Corollary \ref{ZHDFCor1} still true when the hypothesis ``$p$ is a prime number''
is replaced by $\Omega(p) \leq 2$? (see Corollary \ref{main2}.)
Is Corollary \ref{ZHDFCor2} still true when the hypothesis ``one among
$P,Q$ is a singleton''
is replaced by ``$\Card(P) \leq 2$ or $\Card(Q) \leq 2$''? (see Proposition \ref{main1}.)

%--------------------------

\end{document}